\documentstyle{ioplppt}

\input epsf

\begin{document}

\jl{2}
\title{Formation of fundamental structures in Bose-Einstein Condensates }[Fundamental
structures in Bose-Einstein condensates]

\author{T. F. Scott\dag, R. J. Ballagh\dag, and K. Burnett\ddag}
\address{\dag\ Department of Physics,
University of Otago,
P.O.Box 56,
Dunedin,
New Zealand}
\address{\ddag\ Clarendon Laboratory, Department of Physics,
University of Oxford, Parks Road,
Oxford OX1 3PU, United Kingdom.}

\begin{abstract}

The meanfield interaction in a Bose condensate provides a nonlinearity which
can allow stable structures to exist in the meanfield wavefunction. We discuss
a number of examples where condensates, modelled by the one dimensional Gross
Pitaevskii equation, can produce gray solitons and we consider in detail the
case of two identical condensates colliding in a harmonic trap. Solitons are
shown to form from dark interference fringes when the soliton structure,
constrained in a defined manner, has lower energy than the interference
fringe and an analytic expression is given for this condition.

\end{abstract}


\maketitle


The experimental realisation of atomic gas Bose Einstein condensates has
opened the door to the study of quantum phenomena on a mesoscopic scale, and
a major interest is whether qualitatively new phenomena will occur. Some of
the most striking experimental results to date involve bulk dynamics of the
condensate, and the JILA and MIT groups have made quantitative measurements
of collective excitations \cite{excitation}, sound propagation \cite
{AKMDTIK97} and interaction between distinct condensates \cite
{MBGCW97,MAKDTK97}. A beautiful demonstration of quantum interference has
been made by Ketterle's group, where they allowed two distinct condensates
(in the same internal state) to freely expand into each other to produce
fringes \cite{ATMDKK97}. Careful analysis of this experiment \cite{RNSW97}
has shown that the simple picture of linear interference does not adequately
describe the results, and an additional mechanism, collisional self
interaction, plays an important role in the formation of the fringe pattern.
In this paper we describe and analyse a noteworthy new phenomenon arising in
condensates as a result of the self interaction, namely the formation of
fundamental structures.

It is now well accepted that the Gross-Pitaevskii (GP) equation 
\begin{equation}
i\frac{\partial \Psi ({\bf r},\tau )}{\partial {\tau }}=-\nabla ^{2}\Psi (%
{\bf r},\tau )+V({\bf r})\Psi ({\bf r},\tau )+C|\Psi ({\bf r},\tau
)|^{2}\Psi ({\bf r},\tau ),  \label{GPeq}
\end{equation}

provides a good description for the low temperature behaviour of a
condensate in a single internal state. In this equation, (in which
dimensionless units are used as in Ruprecht{\it \ et al}\cite{RHBE95}), $%
\Psi $ can be interpreted as the condensate wavefunction \cite{G97}, $V({\bf %
r})$ is the trap potential, and $C$ is proportional to the number of atoms
in the condensate and the scattering length. The self interaction is
described by the final term of Eq.(\ref{GPeq}), introducing a nonlinearity
into the description which, as is well known from other branches of physics,
can profoundly alter the character of the system behaviour. A notable new
feature that can appear is solitons \cite{Taylor}, which are well known only
in the one-dimensional case, but in the 2 or 3D case, analogous structures
with inherent particle-like robustness, such as vortices, can also occur.
Even in the 1D case, no general results are known for the GP equation.
Morgan et al \cite{MBB97} have shown the existence of solitary wave
behaviour under some well defined conditions. In this paper we present a
simple scenario in which colliding trapped condensates produce a particular
class of fundamental structure, gray solitons. These topological solitons
appear as a density dip with an associated phase kink, and Reinhardt and
Clark \cite{RCpre97}have discussed how the gray solitons resulting from such
collisions might be used to establish the initial phase of the two
condensates. In this paper we focus on the mechanism of formation of the
gray solitons. Using numerical solutions of Eq.(\ref{GPeq}) in one
dimension, we have explored the phenomenon of gray soliton formation and
identified the key mechanisms, thereby allowing us to develop an analytic
description to provide quantitative understanding. The 1D case, although a
considerable simplication over the more realistic 2D or 3D cases,
nevertheless captures some essential characteristics including the coupling
between nonlinearity and spatial variation, and we expect it will provide
guidance for eventual extension into 2 or 3 spatial dimensions. {\sf \ }

We have found that gray solitons can be produced by a variety of means, but
for this paper we have chosen to study in detail the simplest scenario that
enables us to identify and understand the key mechanisms involved, namely
the case of two separated condensates colliding under the influence of a
harmonic trap. This is similar in spirit to the interference experiment of
Ketterle's group, but has some distinctive features which are
important for the analysis we give. As Morgan et al \cite{MBB97} have shown,
a spatially displaced ground state eigenstate of the time independent GP
equation, evolving alone in the trap, retains its shape while executing
simple harmonic motion with period $\tau =2\pi $ in the present units. The
spatial phase profile is then always linear in the spatial variable $x$,
with a slope proportional to the velocity of the condensate. Thus by
choosing an initial state consisting of two ground state eigenstates of
equal particle number, displaced equal and opposite amounts from the center
of the trap \footnote{The total wavefunction is normalised to 1, and the GP
equation evolved with $C$ equal to twice that of each eigenstate.}, we can easily isolate the phenomena that arises
due solely to the collisions.

From an extensive set of simulations, in which we varied the value of $C$,
the initial separation, and the phase difference between the two
condensates, we have found that we can divide the results into two broad
regimes. In the first, which we call the {\it linear regime}, the total
wavefunction $\Psi $ can be very well approximated at any time $\tau $
during the first few periods of oscillation, by linearly superposing the
wavefunctions that each condensate alone would have at $\tau .$ The
condensates, initially well separated, evolve to overlap at the centre of
the trap producing a familiar linear fringe pattern, then separate to
reconstitute themselves into two well separated condensates, regaining their
original shape. The fringe spacing at any time during the overlap is
determined from the $k$ vector (i.e slope of the spatial phase) that the
condensates would have at that time, if oscillating alone in the trap. The
linear regime occurs when the kinetic energy of the condensates during
collision dominates the nonlinear self energy, a condition that requires the
condensates to be initially well separated. An approximate quantitative
criteria for the linear regime can be found by considering the linear
interference pattern that the two condensates would produce at the time of
maximum overlap ($\tau =$ $\pi /2,$ $3\pi /2,$ ...). For an initial
separation of the centres of the condensates of $2d$, then the speed of each
condensate at $\tau =$ $\pi /2$ is $d$ , (in the present units), giving rise
to a fringe spacing of $\pi /d.$ Approximating the separate wavefunctions by
Thomas-Fermi wavefunctions (e.g \cite{EB95}) the condition that the peak
kinetic energy (estimated from the curvature of a $\cos ^{2}$ interference
pattern) exceeds the peak self energy (i.e. at the centre of the fringes),
and hence that the collisional behaviour is linear, takes the form

\begin{equation}
d > \frac{\pi }{2}(3C)^{1/3}.  \label{Cond}
\end{equation}

\vspace{0.5cm} 
\begin{figure}[tbh] 
\centerline{
\epsfxsize=14cm
\epsfbox{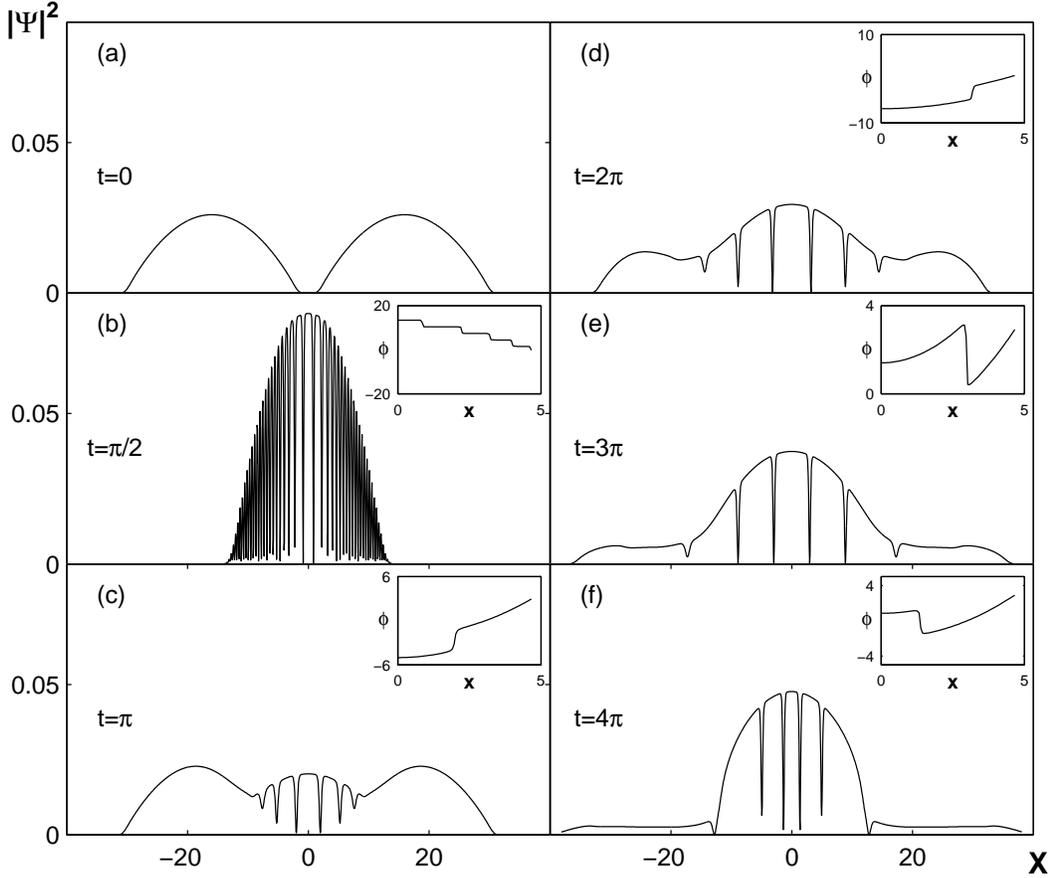}
}
\caption{Evolution of condensate density for two initially
stationary, displaced eigenstates for the case $C = 2000$. Evolution of a
selected region of the phase is shown inset. Initially, phase is flat across
the entire condensate, and $d = 15$.
}
\end{figure} 

When Eq.(\ref{Cond}) is not satisfied, the collisional behaviour falls into
the {\it nonlinear regime}, and a representative example is given in Figure
1, where the total condensate density is shown evolving over two periods of
oscillation. It is clear that the behaviour is not simple linear
superposition. The initial condensates do not reform at the periodic
intervals $\tau =2\pi ,4\pi ,$.. , but rather, the combined condensate
pulsates, contracting and expanding with period $2\pi ,$ and gradually
accumulates at the centre of the trap as time progresses. The most striking
feature however is the appearance of several persistent dark fringes, each
with an associated phase kink of somewhat less than $\pi .$ These dark
fringes remain clearly discernible throughout the course of the simulation,
surviving the extremely complex behaviour that occurs at the time of maximum
contraction ( $\tau =$ $\pi /2,$ $3\pi /2,$ ...). We identify them with gray
solitons, although these have only been fully characterised in the case of an
infinitely extended medium in a homogeneous environment(e.g. see \cite
{Weiner}).%
\begin{figure}[tbh] 
\centerline{
\epsfxsize=7cm
\epsfbox{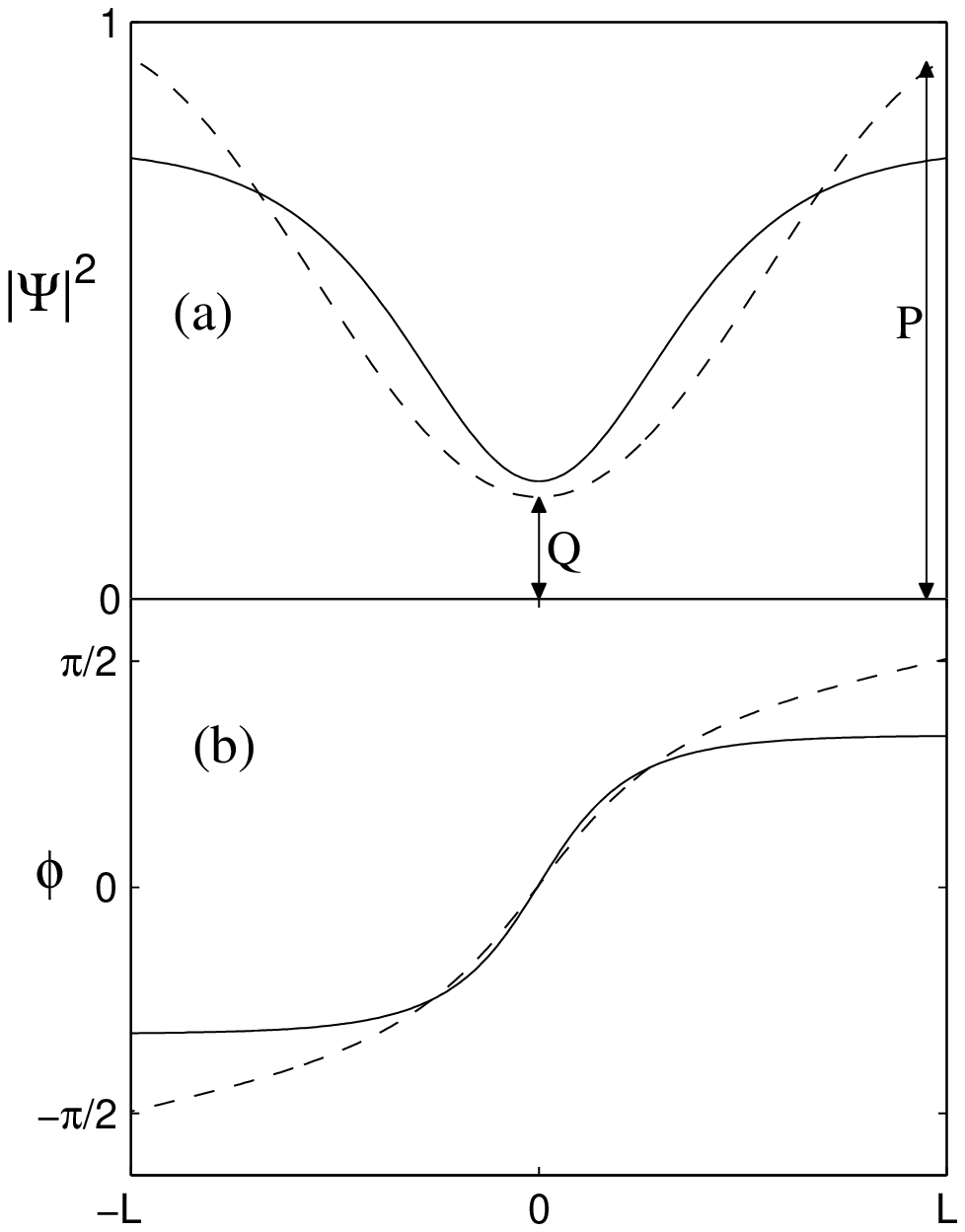}
}
\caption{Comparison of (a)density and (b)phase spatial profiles for a linear
interference fringe (dashed line) and the corresponding gray soliton (solid
line) constrained as described in the text. Here $P = 0.71$, $Q = 0.13$ $C =
2000$, $L = 0.11$ and $\zeta_{f} = 2.7$.
}
\end{figure} 
%
In that case, they can be described analytically, and an example
is shown as the solid line in Figure 2. The depth of the dip characterises
the grayness of the soliton, which is said to be black (or dark) if the wave
density goes to zero. The phase (Fig2(b)) has a corresponding kink in the region of
the density dip, which is a sharp $\pi $ step for a black soliton, but
softens as the soliton becomes gray, thereby giving the gray soliton a
velocity relative to the background density. All of these properties are
possessed by the six dark fringes in Fig1 (c)-(f) , and evolve adiabatically
as the background height changes. We have also shown that the fringes are
very long lived (they are still present when our simulation is extended to $%
\tau =48\pi $) and in other circumstances we have shown that they have a
repulsive interaction and survive collisions with each other. The stability
of the solitons is vitally dependent on the phase kink, without which any
dip quickly evolves to something quite different. The phase kink also
provides a vital key to understanding the formation mechanism of these
prototypical fundamental structures. The formation process begins when an
essentially linear interference pattern starts forming where the two initial
condensates overlap at the centre of the trap. The amplitude is small, so
that nonlinear effects are negligible, and thus the spacing of these first
few fringes is determined by the wavevectors (i.e. velocity) that each
condensate has developed at that time. Fringes offset from the trap centre
arise from the linear superposition of unequal amplitude opposing travelling
waves, and thus have reduced visibility, and an associated phase kink (see
Figure 2). As time progresses, the condensates continue to accelerate into
each other, and as the overlap increases, additional fringes start to
develop further from the center. The central few fringes evolve as local
features, and remain (essentially) locked in place, retaining their initial
spacing while the background density increases about them, but the new outer
fringes are formed with closer spacing, appropriate to the increased
velocity. As we can see from Fig 1 (b), by $\tau =\pi /2$ (which would
correspond to complete overlap in the linear case) a multitude of
interference fringes have formed, and the wider spacing of the central
fringes is evident. Subsequently only the central few develop into gray
solitons. The key underlying physics is that a soliton is a stable structure
while a (linear) interference fringe is not, but an interference fringe will
only develop into a gray soliton if it is energetically favorable to do so.
Otherwise, the interference fringe will disappear on a time scale less than $%
\pi /2.$ To obtain a quantitative understanding we consider the energy of
the two structures. A linear superposition of unequal and oppositely
travelling waves of wavevector $k,$ gives rise to a $\cos ^{2}$ shape which
can be characterised by the height $P$ and $Q$ of the maximum and minimum
density, respectively, as illustrated by the dotted line in Fig 2(a) (dashed
line). The total energy of this wave, in the length $2L$ between fringe
peaks is

\begin{equation}
E_{\sup }=\int_{-L}^{L}\Psi ^{*}(x)H\Psi (x)dx=\frac{CL}{8}\left(
3P^{2}+2PQ+3Q^{2}\right) +k^{2}L(P+Q).  \label{SupEnergy}
\end{equation}
In general a gray soliton is completely specified by two parameters, but if
we now deform this same amount of condensate from the superposition, into a
gray soliton within the same region $-L<x<L$, the constraint on $%
\int_{-L}^{L}\Psi ^{*}(x)\Psi (x)dx$ allows a one parameter family of
solitons to be inserted in the region. This one parameter essentially
determines how much of the central portion of a soliton will appear in the
region, and so if we further require that the soliton has reached a
specified fraction of its asymptotic value at $x=\pm L/2$, then only one
soliton can be drawn. From the standard form of the gray soliton (e.g see
Ref.(\cite{Weiner})) the asymptotic behaviour is determined by $\tanh (\zeta
)\rightarrow 1,$ at large $\zeta $, (where $\zeta \ $is essentially the
spatial coordinate, but renormalised in our case by a function of the
parameters $C,L,P,Q$ ). Choosing a particular value $\zeta _{f}$ to make $%
\tanh (\zeta _{f})=$ $f,$ some fraction close to 1, we may calculate the
energy of this now completely specified soliton to be

\begin{equation}
E_{soliton}=\frac{CL}{4}(P+Q)^{2}+\frac{4\zeta _{f}^{2}(4\zeta _{f}-3)}{%
3CL^{3}}  \label{SolEnergy}
\end{equation}
For example choosing $\tanh (\zeta _{f})=0.99$ gives the soliton whose
density  and phase we show as  solid lines in Fig 2(a) and 2(b) respectively. A
stable structure, the soliton, will form from the linear superposition when
it is energetically favorable to do so, i.e. when $E_{soliton}\leq E_{\sup }.
$ Writing $V$ for the visibility of the superposition (i.e. $V=(P-Q)/(P+Q)$)
the condition of equality $E_{soliton}=E_{\sup }$ , which defines the
boundary of the regime where solitons can form, can be expressed as

\begin{equation}
P=\frac{1+V}{6CV^{2}L^{2}}[\sqrt{9\pi ^{4}+96V^{2}\zeta _{f}^{2}(4\zeta
_{f}-3)}-3\pi ^{2}].  \label{Solcond}
\end{equation}

We have plotted the curve of $P$ versus $V$ in Figure 3 for two values of $%
\zeta _{f}$ with illustrative values of $C=2000$ and $L=\pi /7$. %
\begin{figure}[tbh] 
\centerline{
\epsfxsize=7cm
\epsfbox{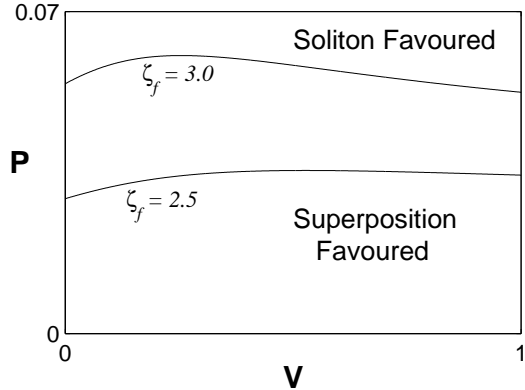}
}
\caption{Regime where linear interference fringe develops into a gray
soliton. The boundary line, Eq.(\protect\ref{Solcond}), is  plotted in terms of
background density $P$ and visibility $V$ of the fringe, for two choices of
$\zeta_{f}$.
}
\end{figure} 
A superposition formed above this boundary line will become a soliton, whereas
a superposition  formed below the line will not convert into the stable
form. The parameters $C$ and $L$ are simply overall multiplicative factors
in Eq.(\ref{Solcond}) and thus move the boundary line up or down without
change in shape. It is clear that increasing the strength of the
nonlinearity (i.e. increasing $C)$ or reducing the speed of condensates at
collision (i.e increasing $L)$ makes it easier for solitons to form. The
choice of $\zeta _{f}$ is important for quantitative purposes, and on
empirical grounds, from considering a range of our simulations, we make the
choice $\zeta _{f}\approx 2.7.$ From Figure 3 we can then see that the
central few fringes of Fig 2(b), (where $L\approx 0.7$ ) are
comfortably in the soliton regime, but as we move to the outer, narrower
fringes of the condensate, $L$ decreases, moving the boundary curve upwards,
and $P$ decreases, with the net result that solitons do not form.

The outcome of a condensate collision, particularly with regard to details
of the solitons that arise, does exhibit some sensitivity to the initial
phase difference of the condensates, as also noted by Reinhardt and Clark.
For example, if the condensates have an initial phase difference of $\pi ,$
an odd number of solitons form, with a dark soliton at the centre of the
pattern, a result easily understood in terms of the formation mechanism
outlined above. There are of course other means to generate dark solitons,
such as collision with a potential barrier, or even by creating a dip in a
condensate eigenstate (for example by temporarily inserting a laser beam, detuned to
provide a repulsive potential). In general the dip will not have the soliton
shape or phase, and is thus unstable and will decompose into a number of
gray solitons which will have a velocity relative to the background,
according to the slope of their phase kink.

Gray solitons can also play an important role in governing the bulk motion of
the condensate. They store energy (predominantly in kinetic form) in a slow
moving structure, and act as a throttle to retard the movement of the
condensate back to the outer region of the trap. This effect can be clearly
seen in Fig. 2(e), where instead of having two well separated condensates
away from the centre, as we would expect in the linear case at $%
t=4\pi ,$ most of the mass has now collected at the center of the trap. We
expect that solitons, (or vortices in the case of 2 or 3 spatial dimensions)
will play an important role in dissipating superfluid flow.

This work was supported financially by the New Zealand Marsden Fund under
contract PVT-603.


\end{document}